# Spectroscopic method for determining the onset of clustering in supersonic gas jets

**Yu. S. Doronin, A. A. Tkachenko, V. L. Vakula, G. V. Kamarchuk**

*B. Verkin Institute for Low Temperature Physics and Engineering of NAS of Ukraine,*
*47 Nauky Ave., Kharkiv, Ukraine*
*E-mail: doronin@ilt.kharkov.ua*

**Abstract**

This article presents an original spectroscopic method for determining the parameters of the initial stage of clustering in supersonic jets. The method was tested on supersonic jets of argon and nitrogen, excited by electrons at a distance of 30 millimetres from the nozzle exit, where the influence of secondary processes on the intensity of the observed radiation is significantly suppressed. For argon, the continuum emission from neutral excited clusters was used as an indicator of cluster formation in the supersonic argon jet. In the case of nitrogen, for which no emissions attributable exclusively to clusters have been recorded, a study was carried out on the effect of clustering on the redistribution of the intensities of atomic lines and molecular bands in the Lyman–Birge–Hopfield system.

## 1. Introduction

Supersonic jets of inert and molecular gases in atomic and cluster flow modes are widely used in a range of scientific and technological applications. The use of atomic supersonic jets as targets for high-energy electron scattering, precision spectroscopy, or laser-plasma experiments enables plasma generation to be spatially separated from substrate interaction, providing precise control in material processing and in the study of complex plasma physics, including shock waves and jet-environment interactions. At the same time, in certain applications of supersonic jets, it is crucial to preserve the jet's atomic composition, since clustering can form undesirable atomic complexes and significantly alter the jet's physical parameters [1–6].

Studies into the interaction of intense laser radiation with cluster supersonic jets [7, 8] have demonstrated greater efficiency in the formation of controlled hot nanoplasma when laser radiation is directed at clusters with small volumes and densities close to those of a solid. Due to the efficient absorption of laser energy within the finite size of the cluster, intense X-ray radiation, a large number of energetic electrons, and highly charged ions are generated, which are followed by a Coulomb explosion or hydrodynamic expansion of the cluster, both of which are of great importance for high-energy physics [9, 10].

Currently, cluster and ion-cluster beams are widely used for the synthesis of thin, atomically smooth films, the doping of ultrathin layers, and the surface treatment of various materials. The high transition energy density generated by individual cluster impacts can also significantly increase the rate of surface chemical reactions, even at low temperatures. A number of published reviews are devoted to the development of cluster technologies and the results achieved through them at the macro- and nanoscale levels [11-13].

For a long time, cluster beams have been the subject of numerous fundamental studies, since the properties of clusters differ from those of the atoms or molecules that comprise them, as well as from those of a continuous medium [14 -16]. Currently, new methods for studying clusters are being proposed and used; for example, the size and density of clusters in a jet can be determined using Rayleigh light scattering [17, 18]. Despite this, there is still no adequate analytical description of clustering in supersonic jets. This is because the condensation process in supersonic jets involves rapid non-equilibrium thermodynamics, complex gas-phase kinetics (cluster nucleation, growth, and fragmentation), and is sensitive to the parameters of the nozzles used and the initial conditions of the flow, which significantly complicates modelling, especially when working with nanoclusters, where classical condensation theory does not work. Experimental diagnostics do not always provide reliable results. Therefore, in many studies, authors continue to use the well-known semi-empirical dimensionless Hagena parameter for condensation, $\Gamma^*$, which allows the average size of the formed clusters to be estimated for a given gas, depending on the initial conditions and the nozzle geometry [19, 20]:

$$\Gamma^* = K_{ch}\left(P_0 \cdot d_{eq}^{0.85} \cdot T_0^{-2.29}\right), \tag{1}$$

where $P_0$ (mbar) is the pressure and $T_0$ (K) is the temperature of the gas at the nozzle inlet, $d_{eq} = 0.74\ d/tan\ \alpha$ for a conical nozzle, $d$ (µm) is the critical diameter, and $a$ (degrees) is the cone angle, $K_{ch}$ is the condensation constant (for Ar, $K_{ch}$ = 1646; for $N_2$, $K_{ch}$ = 528).

In the general case, the average cluster size can be estimated as $N_{cl}$ (at/cl) = 38.4 · $(\Gamma^*/1000)^{1.64}$, although the constant and the exponent may differ for small and large values of $\Gamma^*$ [15]. Furthermore, if $\Gamma^* < 200$, there is no condensation in the supersonic jet; for $200 < \Gamma^* < 1000$, an intermediate stage of jet clustering is characteristic; and $\Gamma^* > 1000$ corresponds to an advanced stage of condensation with the formation of a large number of clusters. It should be noted that, for fixed values of the parameter $\Gamma^*$, there are currently discrepancies of approximately a factor of 2 between experimental results obtained not only by different methods, but also by different authors using the same methodology. These discrepancies can be explained by differences in measurement methods and significant differences in nozzle parameters. Consequently, experiments have been conducted on numerous occasions to determine the limits of applicability of Hagena relations, and methods for refining and correcting them have been proposed [15, 16].

To fully utilise the potential of supersonic jets, it is necessary to know several key parameters. These include: the nature of the jet flow and the density or concentration of atoms, the pressure and temperature of the gas at the nozzle inlet at which the clustering process begins in the jet, the ratio of monomers and clusters in the jet at different stages of clustering, the mass fraction of condensate, and the average cluster size. In [21], the authors used a spectroscopic method based on absolute measurements of the VUV radiation flux generated by the excitation of a supersonic argon jet with an electron beam of 1 keV energy to determine the density of the non-condensed atomic component, the condensate fraction, and the cluster concentration in a supersonic argon jet at given flow parameters.

In this article, we propose using such spectroscopic measurements to determine the parameters at which the clustering process begins in a supersonic jet.

## 2. Experiment

The study was carried out on a spectroscopic setup described in detail in [22]. To generate a supersonic jet, a conical supersonic nozzle was used with a critical cross-sectional diameter of 0.34 mm, a diffuser length of 11 mm, a cone angle of 8.60°, and an exit-to-critical cross-sectional area ratio of 36.7. At a distance of 30 mm from the nozzle outlet, the supersonic jet was excited by an electron beam with an energy of 1 keV and a current of 20 mA. The radiation generated at the point where the supersonic jet intersected with the electron beam was analysed by a vacuum spectrometer attached to the vacuum chamber of the cluster beam generator. A PMT-142 sun-blind photoelectron multiplier, situated behind the exit slit of the monochromator, was used as the radiation detector.

## 3. Results and discussion

### 3.1. Argon

The cathodoluminescence spectra of a clustered supersonic argon jet in the 90–150 nm range contain identical spectral components, the intensity of which varies with changes in the outflow parameters $P_0$ and $T_0$ [22]. Figure 1 shows the spectrum of an argon jet at $P_0$ = 0.1 MPa and $T_0$ = 180 K. The spectrum contains resonance lines emitted by Ar II ions ($\lambda$ = 92.0 nm, $\lambda$ = 93.2 nm) and Ar I atoms ($\lambda$ = 104.8 nm, $\lambda$ = 106.7 nm), as well as the $Ar_2^*$ continuum at $\lambda$ = 127 nm and the $(Ar_4^+)^*$ at $\lambda$ = 137 nm, associated with the emission of localised excitons of the excimer molecule type in vibrationally relaxed states, formed in neutral and ionised clusters. The continuum with a maximum at $\lambda$ = 109 nm (W) is emitted by localised excitons in partially vibrationally relaxed states [21, 22]. The average size of the $N_{cl}$ clusters was determined using a modified empirical Hagena relations [15, 20].

It should be noted that, in the atomic jet flow regime at $P_0$ = 0.1 MPa and $T_0$ > 400 K, only atomic and ionic lines are observed in the spectrum [22].

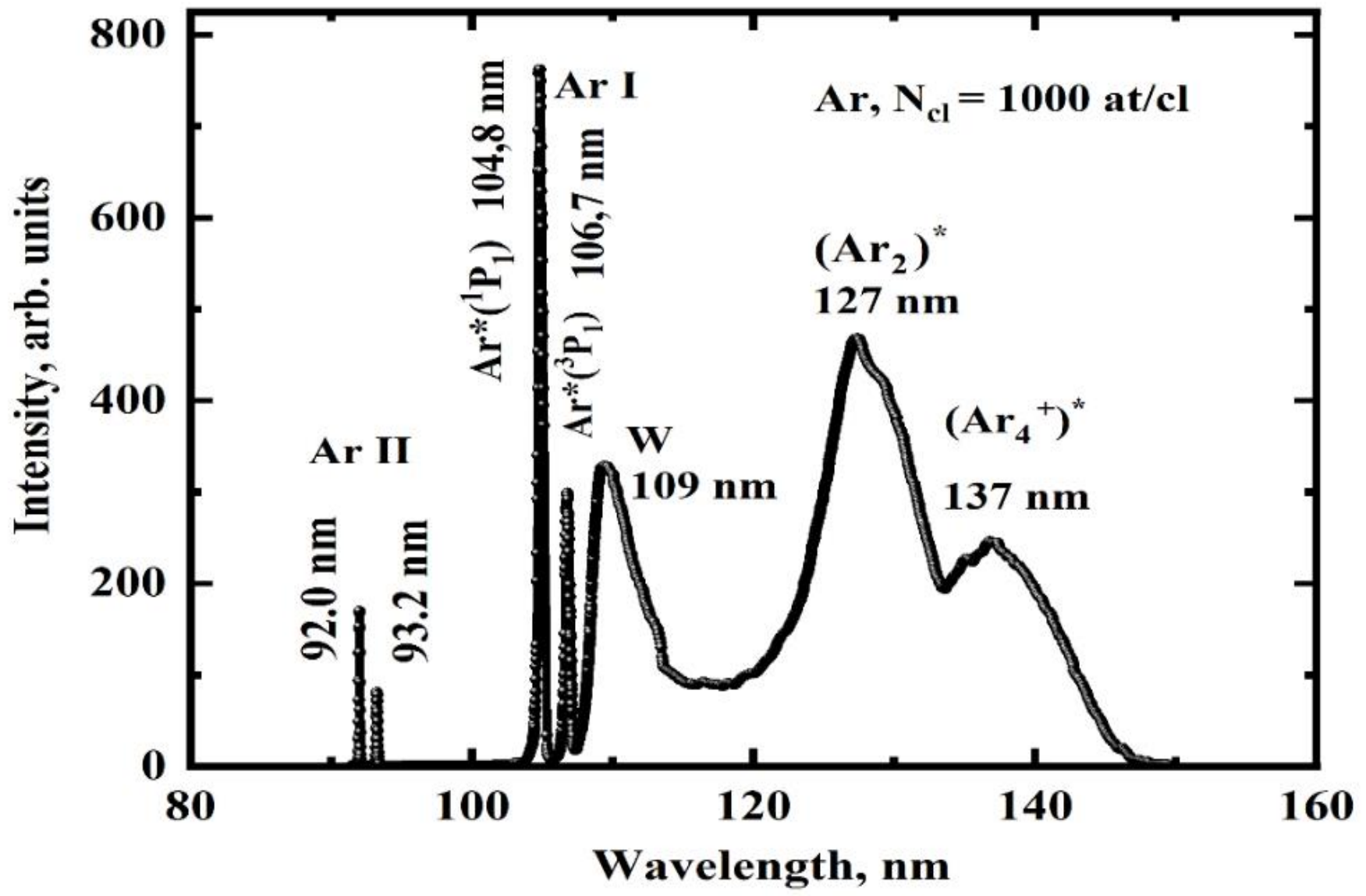

Fig. 1. The emission spectrum of a supersonic argon jet at an electron energy of 1 keV, an electron beam current of 20 mA, $P_0$ = 0.1 MPa and $T_0$ = 180 K.

Experimental studies conducted over a wide range of argon pressures and temperatures at the nozzle inlet have shown that, across all flow regimes, uncondensed atoms are present in the supersonic argon jet [22]. Therefore, an assessment was carried out of the probability of excimer molecule formation in the atomic matrix of a supersonic argon jet. As shown in [21], when using a conical nozzle with similar parameters, the concentration of atoms in the investigated region of the supersonic argon jet varies from 2 to $5 \cdot 10^{15}$ $cm^{-3}$ as $P_0$ and $T_0$ change. In gaseous argon, the dominant pathway for the formation of excimers is the three-particle reaction $Ar^* + Ar + Ar = Ar_2^* + Ar$. The rate of this reaction for argon can be determined using the formula:

$$w = k_3 \cdot [Ar^*] \cdot [Ar]^2 \;, \tag{2}$$

where $w$ is the reaction rate ($cm^{-3} \cdot s^{-1}$), $k_3$ is the rate constant for the three-particle process (for argon, $k_3 \approx 1 \cdot 10^{-32}$ $cm^6 \cdot s^{-1}$ [23]), $[Ar^*]$ is the density of excited argon atoms ($cm^{-3}$), $[Ar]$ is the density of unexcited argon atoms ($cm^{-3}$).

The concentration of excited $Ar^*$ atoms in the region of the jet under investigation is primarily determined by argon atoms in the excited ($^3P_1$ and $^1P_1$) and metastable ($^3P_2$ and $^3P_0$) states of the $3p^54s$ configuration. The region of the supersonic jet flow under analysis is a cylinder with a diameter of 0.3 cm and a length of 1.0 cm. The time taken for the atoms in the supersonic jet to pass through this region is $\tau_{pass} \approx 3 \cdot 10^{-6}$ s. Radiation capture, which is characteristic of resonant radiation [24], increases the lifetimes of the $^3P_1$ and $^1P_1$ states to $\tau_{eff} \approx 10^{-5}$ s, with the result that they become longer than, or comparable to, $\tau_{pass}$. Taking into account the lifetimes of the atoms in these states and the time $\tau_{pass} \approx 3 \cdot 10^{-6}$ s during which they are present in the investigated volume of the supersonic jet, the calculated density of excited argon atoms $[Ar^*]$ is $\approx 2.6 \cdot 10^{12}$ $cm^{-3}$. In that case, the maximum value of the three-particle reaction rate for argon is $w \approx 5.8 \cdot 10^{11}$ $cm^{-3}\, s^{-1}$, whilst the he probability $P_{ex} = \nu \cdot \tau_{pass}$ that a single excited atom will have time to form an excimer during the beam transit time $\tau_{pass}$ can be estimated using the deactivation rate of a single atom $\nu = w / [Ar^*]$. The final value of $P_{ex} \approx 1 \cdot 10^{-6}$ indicates that fewer than 0.0001% of the excited atoms formed in the gas manage to form an excimer. Metastable atoms with lifetimes of 45–56 *s* are carried by the supersonic flow beyond the boundaries of the region under analysis whilst in their initial state and play virtually no part in the reaction under consideration. The analysis carried out confirms that the excimers formed in the gas phase do not contribute to the intensity of the recorded argon continua (see Fig.1). Thus, since the $\lambda$ = 127 nm continuum is a cluster continuum, its dependence on the gas temperature and pressure at the nozzle inlet may serve as an indicator of clustering in the supersonic argon jet.

The dependence of the intensity of the continuum at λ = 127 nm, expressed in relative units, as a function of various pressures $P_0$, is shown in Fig. 2.

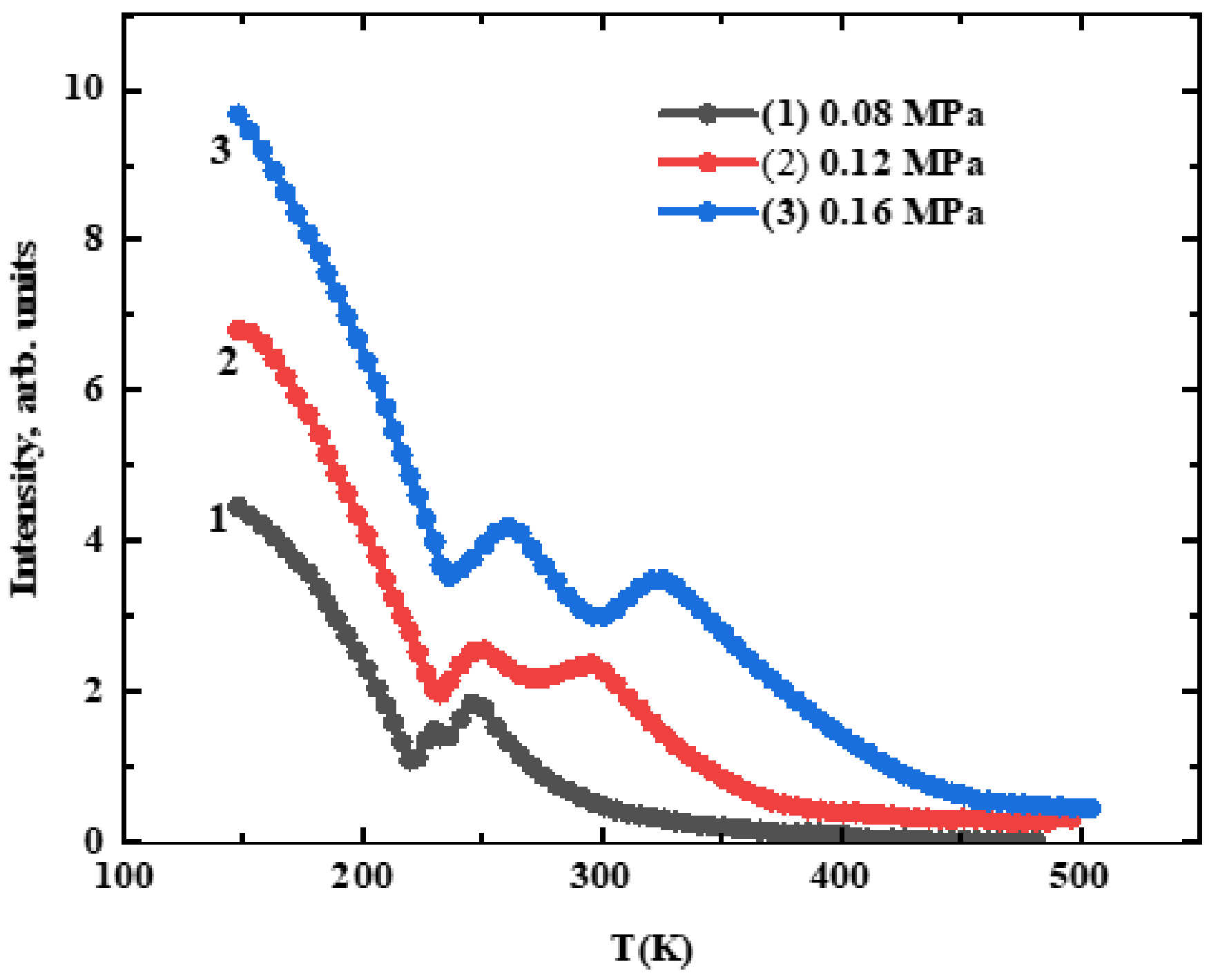

Fig. 2. The temperature dependencies of the intensity of the continuum $\lambda$=127 nm at different pressures $P_0$ of the gas at the nozzle inlet: 1- 0.08 MPa; 2 - 0.12 MPa; 3 - 0.16 MPa.

The figure shows that the emergence and growth of the continuum occur at lower $T_0$ as pressure $P_0$ decreases. In [21] it was shown that the clustering process correlates with a decrease in the intensity of the Ar II line at $\lambda$ = 92.0 nm. Therefore, at the same pressures as for $\lambda$ = 127 nm, temperature dependencies were recorded for the $\lambda$ = 92.0 nm line (Fig. 3).

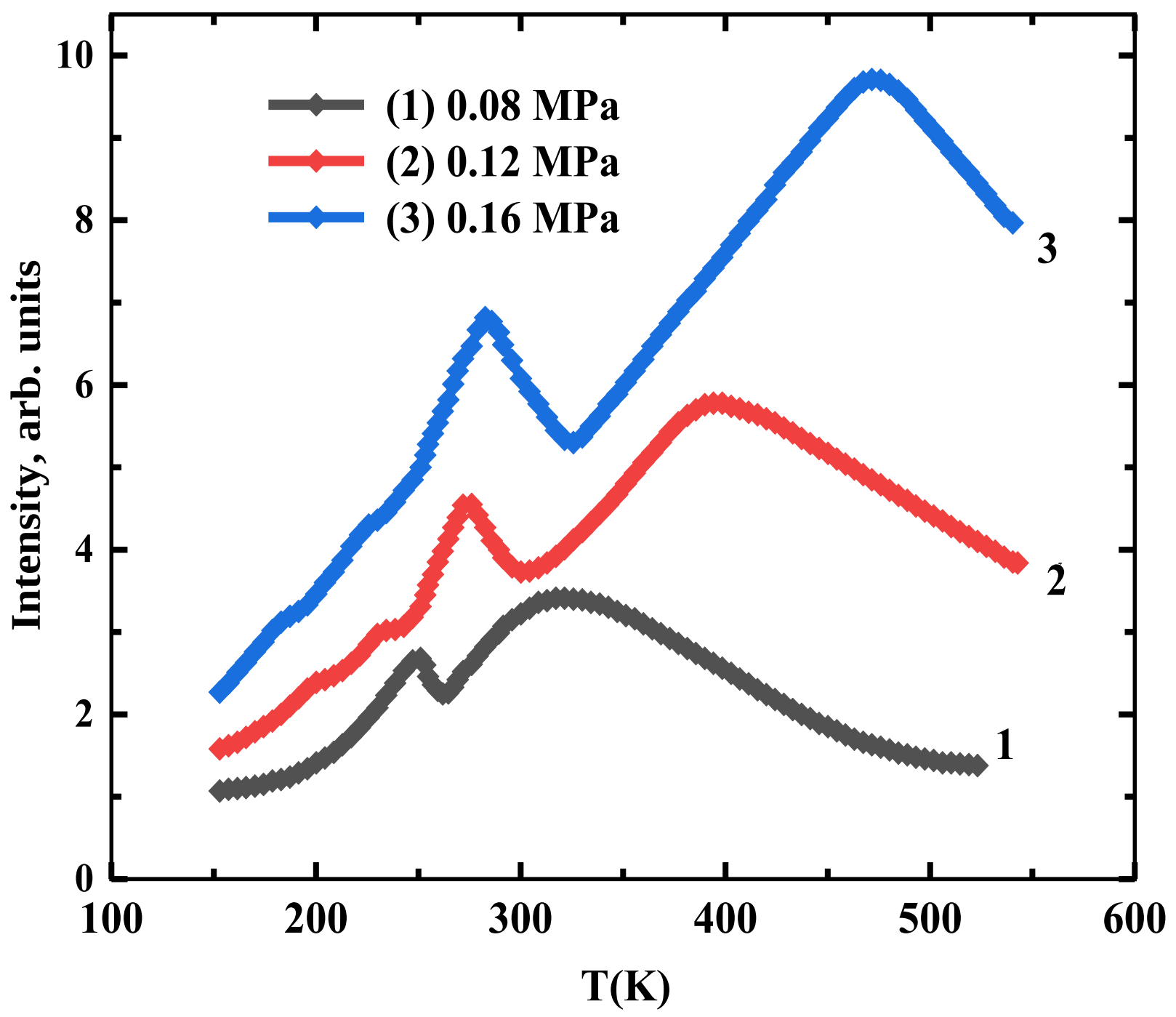

Fig. 3. The temperature dependencies of the intensity of the $\lambda$ = 92 nm line at different pressures $P_0$ of the gas at the nozzle inlet: 1- 0.08 MPa; 2 - 0.12 MPa; 3 - 0.16 MPa.

Analysis of the obtained temperature dependencies *(Fig. 2, 3)* showed that the process of cluster formation in a supersonic argon jet begins at the following values of $P_0$ at the nozzle inlet: $P_0$ = 0.08 MPa – 340 K; $P_0$ = 0.12 MPa – 410 K; $P_0$ = 0.16 MPa – 470 K. The start of the clustering process is characterised by the empirical expression $P_0T_0^{-2.7} \approx 0.011$, where pressure is in Pa, and temperature is in degrees K. The proposed method registers temperature dependencies for lines and continua with an accuracy of approximately 10%.

In accordance with Eq. (1), the following calculated temperature values were obtained for the nozzle used in this study at the specified pressures at which clustering begins ($\Gamma^* > 200$): $P_0$ = 0.08 MPa – 950 K, $P_0$ = 0.12 MPa – 1130 K; $P_0$ = 0.16 MPa – 1280 K. It should be emphasised that the calculated values of $T_0$ at all given $P_0$ significantly exceed the corresponding values of $T_0$ determined experimentally in this work.

The results show that the clustering process in a supersonic argon jet formed by a conical nozzle begins at significantly lower temperatures than those predicted by the condensation scaling parameter $\Gamma^*$, introduced by O. Hagena and co-authors. The exponent in the expression $P_0 \cdot T_0^{-2.7} \approx 0.011$, equal to (-2.7), indicates that cluster formation in a supersonic conical nozzle occurs much faster than in sonic nozzles. In addition, the conical nozzle provides higher gas density and lower temperature in the supersonic expansion region, creating ideal conditions for cluster formation (high supersaturation, high collision frequency).

The observed non-monotonic temperature dependence of the continuum, with a maximum at 127 nm in the 92 nm line as temperature decreases, results from the transition from the initial stage of clusterisation to cluster coalescence (the advanced stage of clusterisation). During this transition, the latent heat of condensation released into the supersonic jet fragments unstable clusters of critical size. Consequently, the density of the jet's atomic matrix and the cluster concentration change, leading to fluctuations in the cathodoluminescence spectra of the supersonic argon.

### 3.2. Nitrogen

Despite extensive experimental studies of the physical properties of nitrogen in various phases across a wide temperature range, there is relatively little data on the clustering of nitrogen in supersonic jets. Experimental studies have mainly focused on cluster structure [25-27], and only a few papers have reported on the spectroscopy of electronic and vibrational transitions in molecular nitrogen clusters [28, 29]. CL spectra of a supersonic nitrogen jet were recorded in the wavelength range 145.0 –175.0 nm, where the well-known Lyman–Birge–Hopfield (LBH) ($a^1\Pi_g - X\,^1\Sigma^+_g$) nitrogen band system is located [30, 31]. The electron beam parameters and experimental geometry were similar to those used in the argon experiments. The spectra were recorded at a nitrogen pressure of $P_0$ = 0.2 MPa and at temperatures of $T_0$ = 270, 200 and 130 K. The nitrogen density in the investigated region of the jet varied between 2 and $5 \cdot 10^{15}$ cm$^{-3}$.

In contrast to argon, the VUV spectra of a supersonic nitrogen jet do not contain any lines or molecular bands that are unambiguously of a cluster nature. Furthermore, the internal degrees of freedom in the $N_2$ molecule give rise to complex energy relaxation pathways that are absent in noble gas clusters. The spectra shown in Fig. 4 reveal atomic nitrogen (NI) lines at 149.3 and 174.3 nm, caused by the electronic transitions $^2P \rightarrow {}^2D_0$ and $^2P \rightarrow {}^2P_0$, which arise as a result of the dissociative excitation of molecular nitrogen. Between these are the most intense molecular nitrogen bands of the Lyman-Birge-Hopfield (LBH) system ($a^1\Pi_g - X\,^1\Sigma^+_g$): 150.1 nm (0–1), 155.5 nm (0–2), 161.1 nm (0–3), 167.2 nm (0–4), 173.6 nm (0–5).

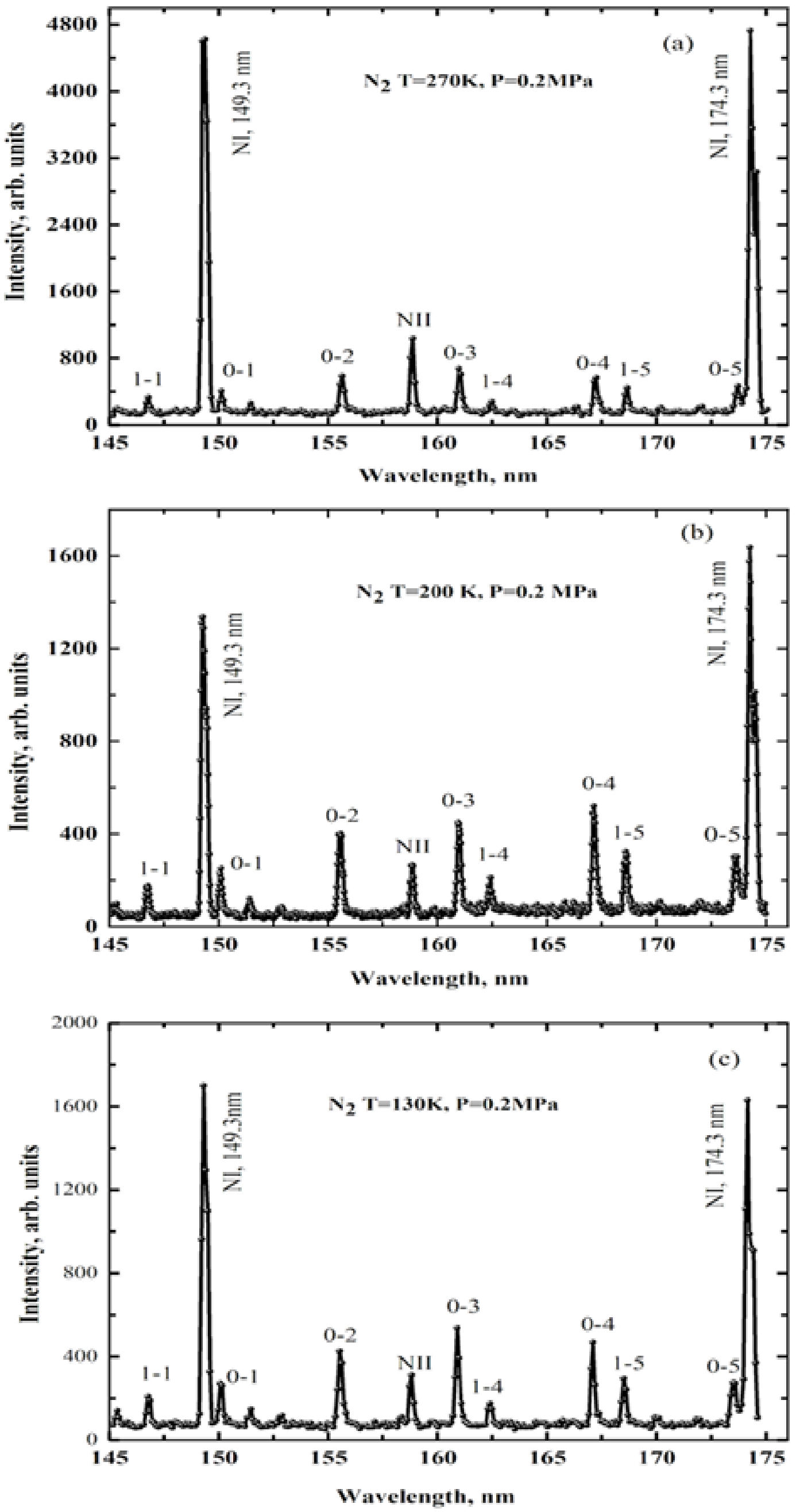

Fig. 4. CL spectra of a supersonic nitrogen jet at $P_0$ = 0.2 MPa and $T_0$ = 270 (a), 200 (b) and 130 K (c). Main features - atomic nitrogen (NI) lines at 149.3 and 174.3 and molecular bands of molecular nitrogen for the LBH system: 150.1 nm (0–1), 155.5 nm (0–2), 161.1 nm (0–3), 167.2 nm (0–4), 173.6 nm (0–5).

Table 1 lists the molecular lines used in the analysis of the CL spectra of a supersonic nitrogen jet. These lines, with wavelengths of 155.5, 161.1 and 167.2 nm, were selected because they are not affected by overlap with neighbouring emission lines. A pronounced non-monotonic variation in the ratio of the intensities of the molecular bands to the atomic line at 149.3 nm (*Im/Ia)* was observed as a function of $T_0$. At $T_o$ = 270 K, the ratio of the integral intensity of the molecular bands to that of the atomic line is close to the corresponding ratio in the gaseous phase of nitrogen [32, 33].

Table 1. Molecular lines with wavelengths of 155.5 (0–2), 161.1 (0–3) and 167.2 (0–4) nm, which were used in the analysis of CL spectra from a supersonic nitrogen jet.

| **Molecular LBH bands (nm)** | **Transition** (v', v") | **270 K** $I_m/I_a$ | **200K** $I_m/I_a$ | **130K** $I_m/I_a$ |
|---|---|---|---|---|
| **155.5** | 0-2 | 0.078 | 0.247 | 0.186 |
| **161.1** | 0-3 | 0.073 | 0.244 | 0.222 |
| **167.2** | 0-4 | 0.049 | 0.281 | 0.184 |

With a further decrease $T_0$ from 270 to 130 K, the intensity of the atomic line decreases by a factor of 3.5, and an increase in the value of $I_m/I_a$ is observed, with a maximum at $T_o$ = 200 K. The observed redistribution of radiation intensity between the atomic and molecular channels can be explained by the competition between direct excitation of nitrogen molecules and dissociative excitation of atoms in the jet in the absence of condensation, and by similar processes at different stages of clustering. The NI atomic line with a wavelength of 149.3 nm, arising predominantly from the dissociative excitation of $N_2$ molecules upon collision with electrons, is proportional to the number of free molecules in the jet. Molecular spectral lines may arise as a result of the direct excitation of nitrogen molecules in the jet, as well as due to excitation relaxation processes within clusters. The number of emitting excited nitrogen atoms in the jet will, in this case, be related to the change in the density of nitrogen molecules in the excitation region resulting from the clustering process, whilst the observed increase in the ratios as the temperature $T_0$ decreases can be explained by the emergence of an additional cluster-associated component of molecular emission in the intensity of the molecular lines.

The maximum value of $I_m/I_a$ at $T_o$ = 200 K is associated with the suppression of the dissociative channel for the formation of excited nitrogen atoms as a result of the initial stage of clustering and the incorporation of nitrogen molecules into clusters. The subsequent slight decrease in the $I_m/I_a$ ratio at $T_o$ = 130 K is the result of a coalescence process, during which the number of free nitrogen molecules in the jet increases due to the decay of unstable condensation nuclei, and, as a consequence, the intensity of the 149.3 nm atomic line increases. Thus, the combination of three independent observations: the strong suppression of atomic emission intensity, variations in the intensity of the LBH molecular lines, and the redistribution of the radiation intensity $I_m/I_a$ between the atomic and molecular channels, points to the influence of clustering on the excitation kinetics of nitrogen long before the appearance of large clusters in the jet, which are reliably detected and analysed using electron diffraction [25, 26].

Calculation of the clustering onset temperature for a supersonic nitrogen jet ($\Gamma^* > 200$) at $P_0$ = 0.2 MPa yields $T_0$ = 860 K according to Eq.1, which significantly exceeds the experimentally determined $T_0 \approx 270$ K for nitrogen. Thus, the calculated threshold criterion $\Gamma^*$ for nitrogen, as for argon, significantly overestimates the temperature at which the clustering process begins under the given parameters of the supersonic jet flowing into a vacuum.

It should be noted that no spectral shifts or broadening were detected in the analysed LBH bands. This suggests that the emitting nitrogen molecules are situated in a local environment that differs significantly from that of bulk solid nitrogen. The nanoscale nature of this phenomenon, including the possible role of surface molecules or locally disordered cluster regions, remains an open question and requires further investigation.

## 4. Conclusions

Based on experimental studies and analysis of the results, it can be concluded that spectroscopy in the extreme ultraviolet (VUV) region can serve as a sensitive method for diagnosing the early stages of cluster formation in supersonic jets. In argon, the onset of clustering is manifested by the appearance and growth of a cluster continuum with a maximum at 127 nm, as well as by a change in the relative intensities of the 93.2 nm ion line. In nitrogen, where no such continuum is observed, clustering manifests as an observable redistribution between atomic and molecular emission channels. It has been established that the calculated threshold criterion $\Gamma^*$ for argon and nitrogen significantly overestimates the temperature (by approximately a factor of three) at which the clustering process begins for the conical nozzle used, under the specified parameters of supersonic jet flow into a vacuum.

Analysis of the obtained temperature dependencies of the intensities $I(T_0)$ of the argon cluster continuum with a maximum at $\lambda$ = 127 nm and the atomic-ion line at $\lambda$ = 92.0 nm has allowed us to establish the empirical relationship $P_0 \cdot T_0^{\ -2.7} \approx 0.\ 011$, which makes it possible to determine the threshold temperature at which the clustering process begins in a supersonic argon jet as the outflow parameters vary.

The proposed spectroscopic method may serve as an additional tool for interpreting the results of studies of supersonic jets of clusters of inert and molecular gases. The experimentally determined boundary parameters $P_0$ and $T_0$ can be used to verify and refine gas-dynamic and kinetic models of condensation in the poorly understood initial region of clustering.

**Acknowledgments**

The research was supported by the National Academy of Sciences of Ukraine project F 9-12.